\documentclass[%
 reprint,
superscriptaddress,
 amsmath,amssymb,
aps
]{revtex4-2}

\usepackage{graphicx}
\usepackage{dcolumn}
\usepackage{bm}
\usepackage{ulem}
\usepackage{color}

\begin{document}

\preprint{APS/123-QED}

\title{Testing gravitational redshift {based} on microwave frequency links onboard China Space Station}

\author{Wenbin Shen}
\email{2891983829@qq.com}
\affiliation{School of Geodesy and Geomatics, Wuhan University, Wuhan 430079, China.}
\affiliation{Time and Frequency Geodesy Center, Wuhan University, Wuhan 430079, China.}
\affiliation{State Key Laboratory of Information Engineering in Surveying, Mapping and Remote Sensing, Wuhan University, Wuhan 430079, China.}

\author{Pengfei Zhang}
\affiliation{School of Geodesy and Geomatics, Wuhan University, Wuhan 430079, China.}
\affiliation{Time and Frequency Geodesy Center, Wuhan University, Wuhan 430079, China.}

\author{Ziyu Shen}
\email{theorhythm@foxmail.com}
\affiliation{School of Resource, Environmental Science and Engineering, Hubei University of Science and Technology, Xianning, Hubei, China.}

\author{Rui Xu}
\affiliation{School of Geodesy and Geomatics, Wuhan University, Wuhan 430079, China.}
\affiliation{Time and Frequency Geodesy Center, Wuhan University, Wuhan 430079, China.}
\author{Xiao Sun}
\affiliation{State Key Laboratory of Information Engineering in Surveying, Mapping and Remote Sensing, Wuhan University, Wuhan 430079, China.}
\author{Mostafa Ashry}
\affiliation{School of Geodesy and Geomatics, Wuhan University, Wuhan 430079, China.}
\affiliation{Time and Frequency Geodesy Center, Wuhan University, Wuhan 430079, China.}
\affiliation{State Key Laboratory of Information Engineering in Surveying, Mapping and Remote Sensing, Wuhan University, Wuhan 430079, China.}
\author{Abdelrahim Ruby}
\affiliation{School of Geodesy and Geomatics, Wuhan University, Wuhan 430079, China.}
\affiliation{Time and Frequency Geodesy Center, Wuhan University, Wuhan 430079, China.}
\affiliation{State Key Laboratory of Information Engineering in Surveying, Mapping and Remote Sensing, Wuhan University, Wuhan 430079, China.}
\author{Wei Xu}
\author{Kuangchao Wu}
\author{Yifan Wu}
\author{An Ning}
\author{Lei Wang}
\author{Lihong Li}
\author{Chenghui Cai}
\affiliation{School of Geodesy and Geomatics, Wuhan University, Wuhan 430079, China.}
\affiliation{Time and Frequency Geodesy Center, Wuhan University, Wuhan 430079, China.}

\date{\today}

\begin{abstract}
    In 2022 China Space Station (CSS) will be equipped with atomic clocks and optical clocks with stabilities of $2 \times 10^{-16}$ and $8 \times 10^{-18}$, respectively, which provides an excellent opportunity to test gravitational redshift {(GRS)} with
    higher accuracy than previous results.
    Based on high-precise frequency links between CSS and a ground station, we formulated a model and provided simulation experiments to test {GRS}.
    Simulation results suggest that this method could test the GR at the accuracy level of $(0.27 \pm 2.15) \times10^{-7}$, more than two orders in magnitude higher than the result of the experiment of a hydrogen clock on board a flying rocket more than 40 years ago.
\end{abstract}

\maketitle


\section{Introduction}\label{I}
Scientists paid great attention to testing the gravitational redshift {(GRS)} effect, one of three classic predictions of general relativity theory (GRT).
In 1976 scientists conducted the Gravity Probe-A (GP-A) experiment mission to test {GRS}  \cite{Vessot1979}.
A hydrogen atomic clock was on board a rocket in this mission, which flew about two hours in space.
Based on the microwave links between the rocket and several ground stations, the GP-A experiment has tested the {GRS} at the accuracy level of $7 \times 10^{-5}$.
To improve the accuracy level, one should consider two aspects.
In one aspect, we try to use better clocks with higher stability and accuracy.
In another aspect, we try to create a condition that the gravitational potential difference between two points is as significant as possible.
For instance, using optic clocks with stabilities around $10^{-18}$, Katori's group \cite{Takamoto2020-rb} measured a height of 450 m with an accuracy of 4.1 cm, by which the accuracy of testing {GRS}  achieves $9.1 \times 10^{-5}$, a little worse than the previous rocket-experiment result of $7 \times 10^{-5}$~\cite{Vessot1980}.
Another example, based on observations from two eccentric-orbit Galileo satellites E14 and E18 (the stabilities of the onboard clocks are around $10^{-15}$/day), several groups tested {GRS} with an accuracy level around $2.5\sim 4.5 \times 10^{-5}$~\cite{Kouba2021,Delva2018,Herrmann2018}.

{Recently, a few space atomic clock projects have been put forward. For example, the Atomic clock Ensemble in Space (ACES) mission onboard the international space station (ISS) and the China Space Station (CSS) mission. The ACES mission will be equipped with an atomic clock with long-term stability \ $2\times 10^{-16}$~\cite{cacciapuoti2011,meynadier2018}, and use its two independent time and frequency transfer links, {including} three microwave links (MWL) and two European Laser Timing (ELT) optical links to distribute time and frequency scale to the ground stations. By using one uplink and two downlinks MWL, a tri-frequency combination method for time-frequency transfer is constructed according to the accuracy of the ACES atomic clock~\cite{blanchet2001,linet2002}. Verified by simulation experiments the tri-frequency combination method may test GRS at a level of  $10^{-6}$~\cite{Sun2021,Shen2021}. The CSS and ACES are different in design. The CSS {will be equipped with} an optical atomic clock whose long-term stability is {$8\times 10^{-18}$~\cite{Guo2021,Wang2021,Sun2021,Shen2021}, and two up-microwave links from a ground station to CSS and two down-microwave links from CSS to the ground station will be extablished. One uplink signal and one downlink signal have the same frequency of 30.4 GHz but different circular polarization directions}. Considering the MWL characters, the errors caused by the propagation will be eliminated. This study created a new frequency transfer model used in the space-ground frequency transfer to test GRS, which is at least an order of magnitude higher than the {results given by previous studies}.}
	
\section{China Space Station}\label{II}
{The CSS designed three modules, one core module (CM) and two experimejnts modules (EM). } The CSS has recently been launched to an orbit with a height around 400 km above the ground since April 29, 2021. From June 17 to September 16, 2021, three Chinese astronauts stayed in the TianHe core module of CSS to prepare for future experiments. Since October 16, another three Chinese astronauts have been successively launched in the core module for further preparatory work. {The CSS will stay in orbits for at least ten years. {For the purpose of time and frequency applications, three or more ground stations have been or will be established, including for instance Beijing, Xi’An, Shanghai, and Wuhan stations. In addition, it will be extablished the space-ground signals links between CSS and ground stations, including microwave links and laser links}.} 

{Experiments module II will be equipped with a set of clocks attached to the core module moving in orbit in September 2022.} The set of clocks consists of a hydrogen clock with daily stability of $2 \times 10^{-15}$, a cold-atom microwave fountain clock with daily stability of $2 \times 10^{-16}$, and an optical atomic clock with long-term stability of $8\times 10^{-18}$~\cite{Guo2021,Wang2021,Sun2021,Shen2021}. {There will be four microwave links between CSS and a ground station of interest for time and frequency transfer.} The MWL consist of two uplinks and two downlinks as listed in Table~\ref{tab:table1}. {Compared with the ACES mission, there are two changes. One is that the CSS mission has an additional uplink with its frequency equal to the frequency of one of the downlink signal. Another is that the uplink and downlink signals are left-hand and right-hand circularly polarized, respectively. {In ACES, there are three microwave signals without polarization, hence it is designed that the three signals have} different frequencies to distinguish the up and downlinks. While, for the CSS mission when a microwave signal is polarized, incident upon an antenna from a given direction, it will be received by the antenna with same polarization character (say both right-hand circularly polarized). This technology is {referred to as} polarization match. The antenna with corresponding polarization character is selected according to different polarization natures of the microwaves to distinguish the uplink and downlink signals~\cite{IEEE1983}.}

\begin{table*}[htbp]]
    \caption{
        \label{tab:table1}
        Parameters of China Space Station (CSS) and frequency links.
    }
    \begin{ruledtabular}
        \begin{tabular}{ll}
            \textbf{Parameter}              & \textbf{Value}                                            \\ \hline
            Altitude                        & $\sim$400 km                                              \\
            Uplink frequency                & 30.4 GHz, 26.8 GHz (both left-hand circularly polarized)  \\
            Downlink frequency              & 30.4 GHz, 20.8 GHz (both right-hand circularly polarized) \\
            Orbit inclination               & $41.5^\circ$                                              \\
            Minimum visible elevation angle & $5^\circ$                                                 \\
            Observation cutoff elevation angle & $15^\circ$                                                 \\
        \end{tabular}
    \end{ruledtabular}
\end{table*}

\begin{figure}[htbp]
    \includegraphics[width=8.6cm,keepaspectratio]{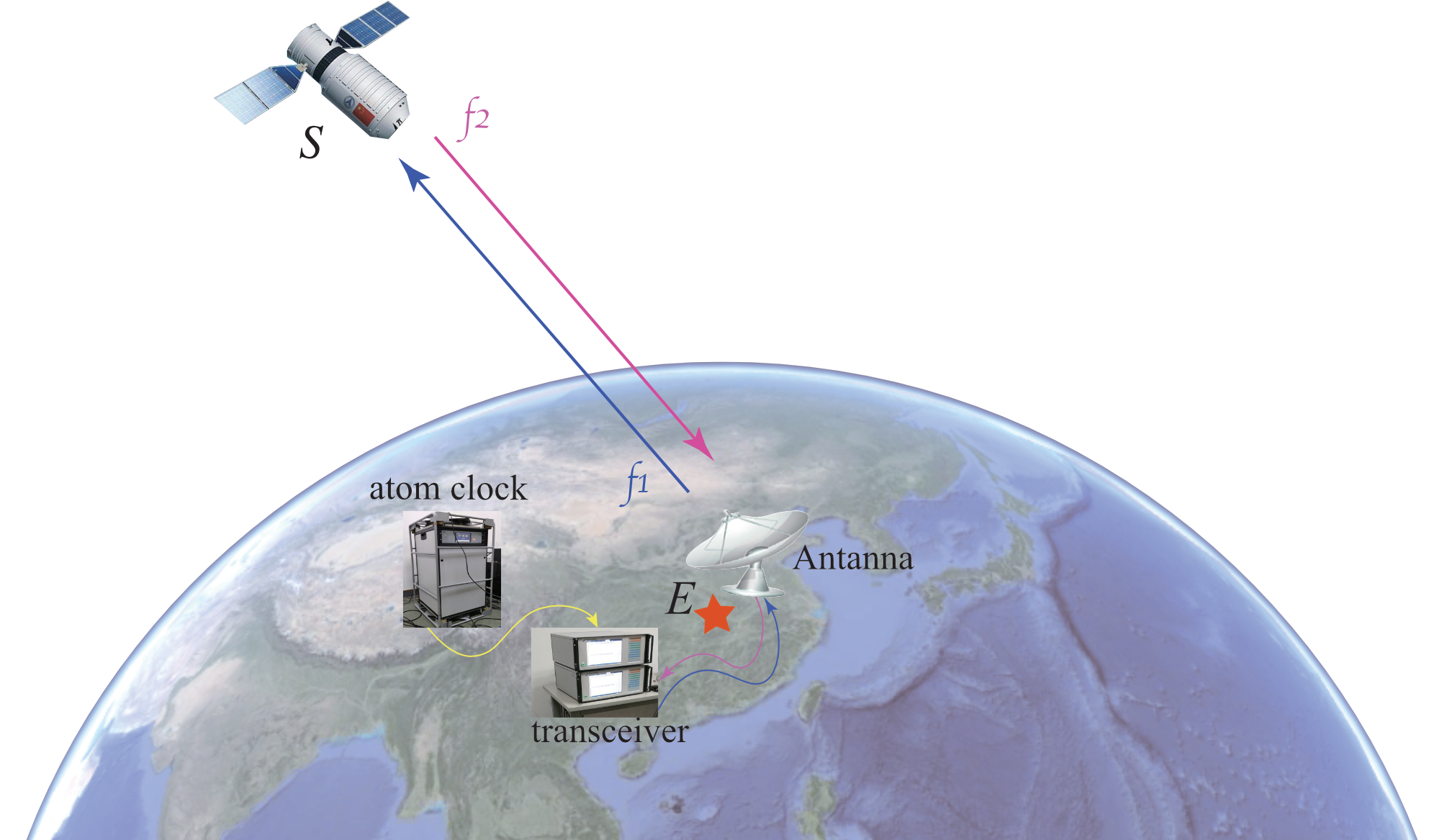}
    \caption{\label{fig1}
        One up-link and one down-link of microwave signals between the China Space Station (CSS) noted as S and a ground station (E) at Wuhan, China.
        The up-link ($f_1$) and down-link ($f_2$) are respectively left- and right-hand circularly polarized signals both with the frequency 
        30.4 GHz. The antenna and the atomic clock at ground are linked by a transceiver. The received signals from CSS are compared with the local signals generated by the ground clock.  
    }
\end{figure}

Here we use only the uplink and downlink signals both with frequency 30.4 GHz (see Fig. \ref{fig1}), and other two links with different frequencies (see Table
\ref{tab:table1}) will not be used in this study. {{Sicne the frequencies of the uplink and downlink are the same, the uninterested frequency shifts caused by the Dopper effects, troposphere and ionosphere effects are the same. Making a difference between} the up and downlinks, we may cancel the {uninterested frequency shifts} and double the GRS, which provides an opportunity to test GR with higher accuracy.} 

\section{Relativistic model for frequency transfer of the CSS}\label{III}
The up-link and down-link signals between CSS (denoted as $S$) and ground station (denoted as $E$) satisfy the following expression:
\begin{eqnarray}
    f_{IJ}^r = &f_{IJ}^e - \Delta {U_{IJ}} + \Delta {f_{{\rm{Dop }}1 - IJ}} + \Delta {f_{{\rm{Dop }}2 - IJ}} \nonumber\\
    &+ \Delta {f_{{\rm{tro - IJ }}}} + \left( {\Delta {f_{ion1 - IJ}} + \Delta {f_{{\rm{ion }}2 - IJ}}} \right) \nonumber\\
    &- \Delta {f_{ti - IJ}} - \Delta {f_{pl - IJ}} + {\varepsilon _{IJ}}
    \label{eq:one}
\end{eqnarray}
where $I, J = E,S, I\neq J$, $E$ and $S$ denote Earth and space station, respectively; $f_{IJ}^r $ is the recieved frequency of the signal at $J$ coming from $I$, $f_{IJ}^e$ is the emitting frequency of the signal at $I$ toward $J$; $\Delta {U_{IJ}} = {U_J} - {U_I}$, ${U_I}$ is the gravitational potential at $I$, $\Delta {f_{{\rm{Dop }}1 - IJ}}$ and $\Delta {f_{{\rm{Dop }}2 - IJ}}$ are the first and second order Doppler frequency shifts of the signals transmitting from $I$ to $J$, $\Delta {f_{{\rm{tro - IJ }}}}$ is the frequency shift of the signal transmitting from $I$ to $J$ caused by the troposphere, $\Delta {f_{ion1 - IJ}}$ and $\Delta {f_{{\rm{ion }}2 - IJ}}$ are the first- and second-order frequency shifts of the signal transmitting from $I$ to $J$ caused by the ionosphere, $\Delta {f_{ti - IJ}}$ and $\Delta {f_{pl - IJ}}$ are frequency shifts caused by tides and external celestial bodies (including the sun, moon, Mercury, Venus, Mars, Jupiter, etc.), respectively, ${\varepsilon _{IJ}}$ are the un-modeled errors of the signals transmitting from $I$ to $J$, including for instance the clock errors, random errors and other noises. As presently designed in CSS, the uplink signal with frequency ${f_1} = 30.4$ GHz is left-hand circularly polarized, and the downlink signal with frequency ${f_2} = 30.4$ GHz is right-hand circularly polarized \cite{Guo2021,Wang2021}. 
Hence, we have $\Delta {f_{{\rm{ion }}2 - ES}} = \Delta {f_{{\rm{ion }}2 - SE}}$ \cite{Hoque2012}. In addition, taking into account the following relationships, $\Delta {U_{IJ}} =  - \Delta {U_{JI}}$, $\Delta {f_{{\rm{Dop }}1 - IJ}} = \Delta {f_{{\rm{Dop }}1 - JI}}$, $\Delta {f_{Dop2 - IJ}} = \Delta {f_{{\rm{Dop }}2 - JI}}$, $\Delta {f_{{\rm{tro - IJ }}}} = \Delta {f_{{\rm{tro - JI }}}}$, $\Delta {f_{{\rm{ion }}1 - IJ}} = \Delta {f_{{\rm{ion }}1 - JI}}$, $\Delta {f_{ti - IJ}} =  - \Delta {f_{ti - JI}}$, and $\Delta {f_{pl - IJ}} =  - \Delta {f_{pl - JI}}$, from Eq.~(\ref{eq:one}), subtracting $f_{ES}^r$ from $f_{SE}^r$, we obtain
\begin{eqnarray}
    \Delta {U_{SE}} = \frac{{f_{SE}^r - f_{ES}^r}}{2} -  \Delta {f_{ti - SE}} - \Delta {f_{pl - SE}} + \varepsilon
    \label{eq:two}
\end{eqnarray}
where $\varepsilon  =  - \left( {{\varepsilon _{SE}} - {\varepsilon _{ES}}} \right)/2$. 
On the right-hand side of Eq.~(\ref{eq:two}), the first term is directly observed. 
The second and third terms, $\Delta {f_{ti - SE}}$ and $\Delta {f_{pl - SE}}$, can be calculated by relevant models. 
For instance, the tides can be corrected by the software ETERNA \cite{Schuller2020,Hartmann1995,Wenzel1997}, and general formulas can directly correct the influences by external celestial bodies \cite{Hoffmann1961,Kleppner1970-db}.  
In Eq. (\ref{eq:two}), both the first- and second-order ionospheric frequency shifts vanish because one left-hand (up-link) and another right-hand (down-link) circularly polarized wave signals are designed in CSS. In practice, since Eq.(\ref{eq:two}) is not rigorous, and the uplink signal's path does not coincide with the downlink signal's path, after the combination of the up and down frequency signals, there are still following residual errors: first- and second-order Doppler frequency shifts, troposphere frequency shift, ionosphere frequency shift, as shown in Table \ref{tab:table2}.

At a given time $t$, we can obtain an observation of $\Delta {U_{ES}}$.
Comparing the observation with the corresponding true (model) value $\Delta {V_{ES}}$, we can get the offset between the observed value and the model value, expressed in relative difference as
\begin{eqnarray}
    \alpha = \frac{\Delta U{(t)_{ES}}-\Delta V{(t)_{ES}}}{V{(t)_{ES}}}
    \label{eq:four}
\end{eqnarray}
which discribes the deviation of the observed result based on GRT from the true (model) value. If GRT holds, $\alpha =0$.

\section{simulation experiments}\label{IV}
In our simulation experiments, we chose Luojia time and frequency station (LTFS) at Wuhan as the ground station, the coordinate of which is ($30^\circ 31^\prime 51.90274^{\prime \prime}$ N, $114^\circ 21^\prime 25.83516^{\prime \prime}$ E, 25.728 m). The CSS parameters and signals' original emitted frequency values are shown in Table~\ref{tab:table1}. The observations in our simulation experiments are the received frequency signals between the
ground station LTFS and CSS, denoted as $f_{IJ}^r$, which includes various frequency shifts caused
by different factors, as expressed by Eq.~(\ref{eq:one}). Here we simulate the first-order Doppler frequency shifts, relativistic frequency shifts {(including gravitational redshift and second-order Doppler effect), atmospheric frequency shift, tidal frequency shift, and clock noises to obtain the received frequency values.

In the simulation experiments, we set the stability of the onboard optical clock of CSS as $2 \times 10^{-15}$/s, and its long-term stability is about $8 \times 10^{-18}$.
The other effects that appeared in Eq.~(\ref{eq:one}) are simulated by the orbit parameters data, which can be downloaded from http://www.celestrak.com/NORAD/elements/stations.txt, including the position, velocity, and acceleration information about the CSS. To obtain a better result, we set the observation cutoff elevation angle of the CSS larger than $15^\circ$. In the simulation, the accuracies of radial orbit position and velocity are set as 10 cm and 1 mm/s, respectively \cite{Sun2021,Shen2021}.

In the simulation of the atmospheric effects on the frequency, we select the Saastamoinen model
\cite{Saastamoinen1972-de}, VMF1(Vienna Mapping Function) \cite{Boehm2013}, and meteorological parameters in the space region near and around the LTFS to simulate the tropospheric frequency shifts. We select IRI (International Reference Ionosphere) model \cite{Bilitza1993} and IGRF-13 (international geomagnetic reference field) model \cite{Baerenzung2020} to simulate the first- and second-order ionospheric frequency shifts. We use the IERS (International Earth Rotation and Reference Systems) Convention 2010 \cite{Petit2010} to calculate the tidal harmonic parameters and consequently simulate the frequency shifts caused by the tidal effect.

According to the simulation setup, we may obtain the observed values $f_{IJ}^r$ based on emitting frequency values.
By Eq.~(\ref{eq:two}), the value of $\Delta {U_{SE}}$ could be calculated.
We use EGM2008 model \cite{Pavlis2008} to calculate the gravitational potentials at CSS and ground station to obtain $\Delta {V_{ES}}$ as the true value.


We have conducted simulation experiments to estimate the feasibility of a {GRS} test based on CSS. The experiment lasts for one day, from May 22 00:00 to May 23 00:00, 2021. {The orbit data were generated by a Two-line Element Set (TLE)~\cite{Li2012}, which is used to describe a spacecraft orbit
	and to predict its position and velocity.} The trace of CSS during one day was depicted in Fig.~\ref{fig2}. The gravitational potential at the space station was calculated by EGM2008~\cite{Pavlis2012}. The precision of EGM2008 is about several centimeters at the CSS' orbit. The ground station LTFS is located at Wuhan, China, and its gravitational potential can be determined by leveling at the precision of centimeter-level or better. Suppose $\alpha = 0$, and then we can calculate the true {GRS} value (model value) based on the general relativity theory (GRT).

\begin{figure}[htbp]
    \includegraphics[width=8.6cm,keepaspectratio]{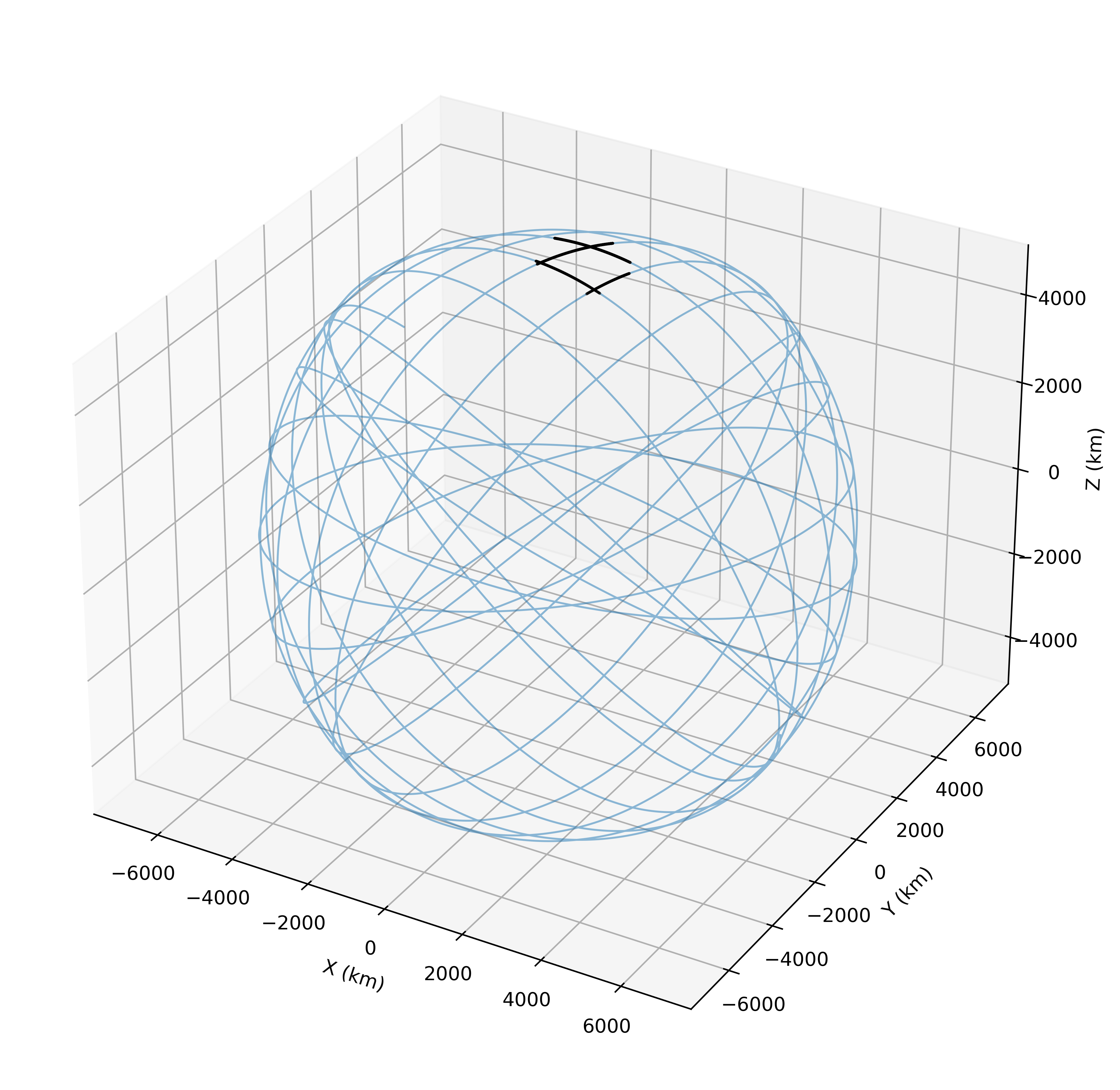}
    \caption{
        The orbit of CSS in Earth Centered Earth Fix (ECEF) coordinates during the observations in one day.
        The ground station is located at Wuhan, China ($30^\circ 31^\prime 51.90274^{\prime \prime}$ N,
        $114^\circ 21^\prime 25.83516^{\prime \prime}$ E, 25.728 m). The observation period is from June 1, 00:00 to June 2, 00:00, 2021.
        The light-colored dash lines denote when the CSS and CSS and Wuhan station are not intervisible.
        The solid black lines denote when CSS and Wuhan station are intervisible and observation values are available.
    }
    \label{fig2}
\end{figure}

The next step is considering various error sources and simulating the observation values for a real case.
For example, one of the primary error sources in the {GRS} test is clock instability.
In our experiment the stability of on-board clock is set as $2 \times 10^{-15}/\sqrt\tau$ ($\tau$ in second), same stability as the designed by CSS mission.
The stability of ground clock is supposed to be better, reaching $1 \times 10^{-15}/\sqrt\tau$.

{According to Allan~\cite{Allan1991}, the clock noises {are classified as five different} types of stochastic noises, including Random Walk Frequency Modulation (RWFM), Flicker Frequency Modulation (FFM), White Frequency Modulation(WFM), Flicker Phase Modulation (FPM) and White Phase Modulation (WPM). We analyzed the ratios of the five kinds of noises to the clock signals and restore the {clock signals according to the  different ratios}. {We set the stabilities of ground clock and on-board clock as $1.0 \times 10^{-15} / \sqrt{\tau}$ and $2.0 \times 10^{-15} / \sqrt{\tau}$, respectivily (see Table~\ref{tab:table2}). Then we use a python library Allantools (https://github.com/aewallin/allantools) to} generate the series of clock frequency errors based on the characteristics of atomic {clocks. Their} total Allan deviations~\cite{Allan1991} are depicted in Fig.~\ref{fig3}.}
\begin{figure}[htbp]
    \centering
    \includegraphics[width=8.6cm,keepaspectratio]{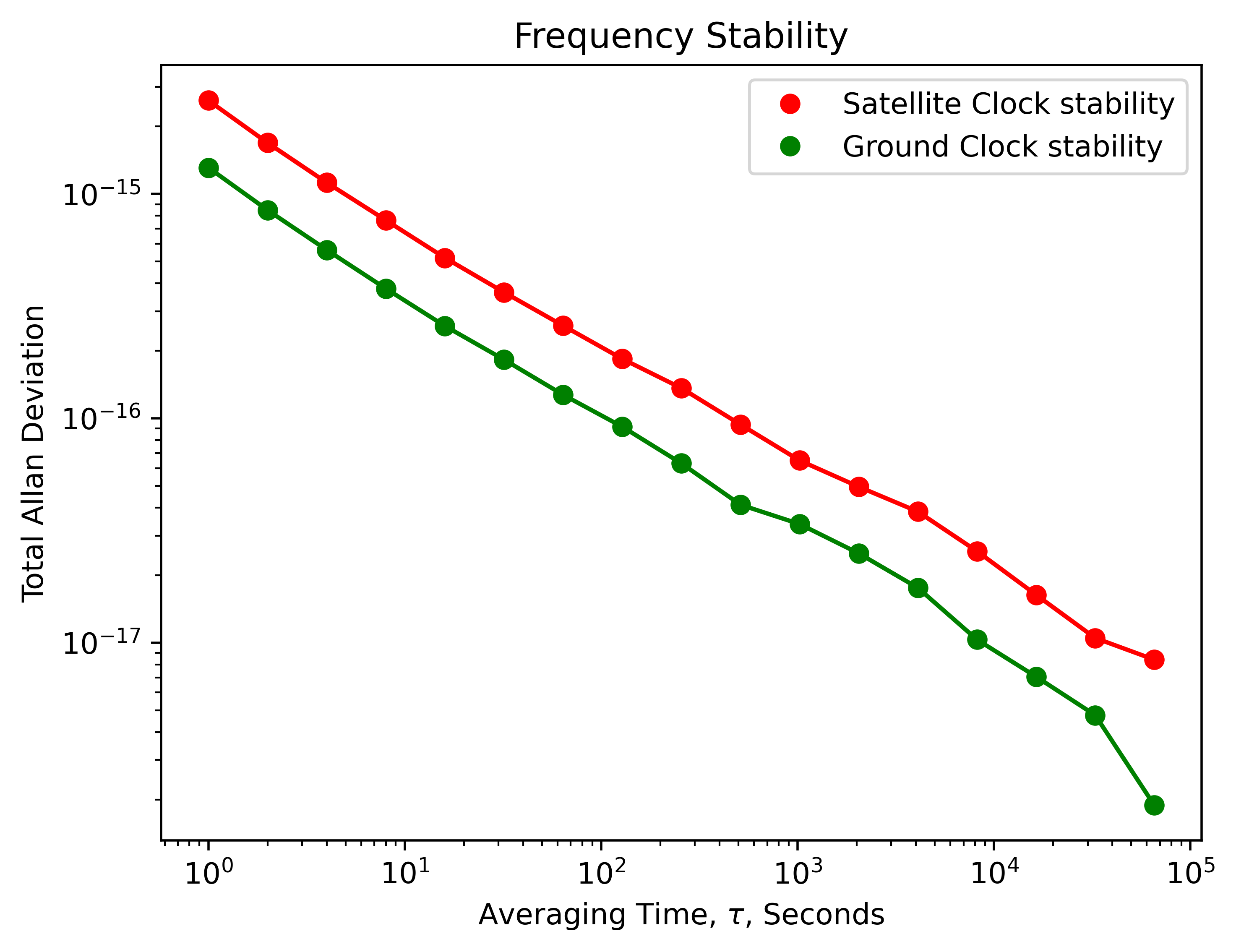}
    \caption{
        The total Allan deviation of the clocks' frequency stability.
        The stability of on-board clock is about $2.0 \times 10^{-15}/\sqrt\tau$ ($\tau$ in second), and the stability of ground clock is about $1.0 \times 10^{-15}/\sqrt\tau$.
    }
    \label{fig3}
\end{figure}
Another important influence factor is the frequency shift caused by the atmosphere.
The local weather station can record the meteorological elements of the ground station.
Then, the signal’s tropospheric frequency shift can be estimated based on the GPT2-mapping function.
The ionospheric frequency shift can be calculated according to International Reference Ionosphere (IRI) model \cite{Bilitza2017}.
The precision of the atmosphere model is regarded as residual errors and added to the frequency shift observation.
Other error sources such as the orbit error of CSS, the EGM2008 error, tidal correction residual are also considered.
Each of them is simulated as Gaussian errors plus a randomly set systematic offset.
The magnitudes of error sources are listed in Table~\ref{tab:table2}.
\begin{table*}[htbp]
    \caption{
        \label{tab:table2}
        Error sources and their magnitudes in the simulation experiments.}
    \begin{ruledtabular}
        \begin{tabular}{ll}
            \textbf{Error sources}                 & \textbf{Magnitudes}                                                                          \\ \hline
            On-board clock stability               & $2.0 \times 10^{-15}/\sqrt\tau$                                                              \\
            Ground clock stability                 & $1.0 \times 10^{-15}/\sqrt\tau$                                                              \\
            Troposphere influence residual         & 5\% of the  tropospheric frequency shift after combination$^*$                                                    \\
            Ionosphere influence residual          & 10\% of the ionospheric frequency shift after combination$^*$ \\
            Position accuracy of CSS               & $\pm$0.1 m                                                                                   \\
            Velocity accuracy of CSS               & $\pm 1.0 \times 10^{-3}$m/s                                                                  \\
            Gravitational potential model accuracy & $\pm$0.3 $m^2/s^2$ (CSS), $\pm$0.1 $m^2/s^2$ (ground)                                        \\
            Tide influence residual                & 0.1 $m^2/s^2$                                                                                \\
        \end{tabular}
        \noindent $*$: Combinantion of up and down frequency signals, exprssed as eq.(\ref{eq:two}) 
    \end{ruledtabular}
\end{table*}

Once the simulated observation frequency values are obtained, they are compared with the true
frequency shift values and part of the offset between them (a short period lasts for about 5 minutes) are depicted in Fig.~\ref{fig4}. We can see that the most prominent offset component comes from clock error, which is at least two magnitudes larger than other error influences. The influence of ionosphere frequency shift shows apparent fluctuation because the total electron content (TEC) distribution varies at different latitudes, and the extent of path discrepancy of uplink and downlink signal also varies at different elevating angles.

\begin{figure}[htbp]
    \includegraphics[width=8.6cm,keepaspectratio]{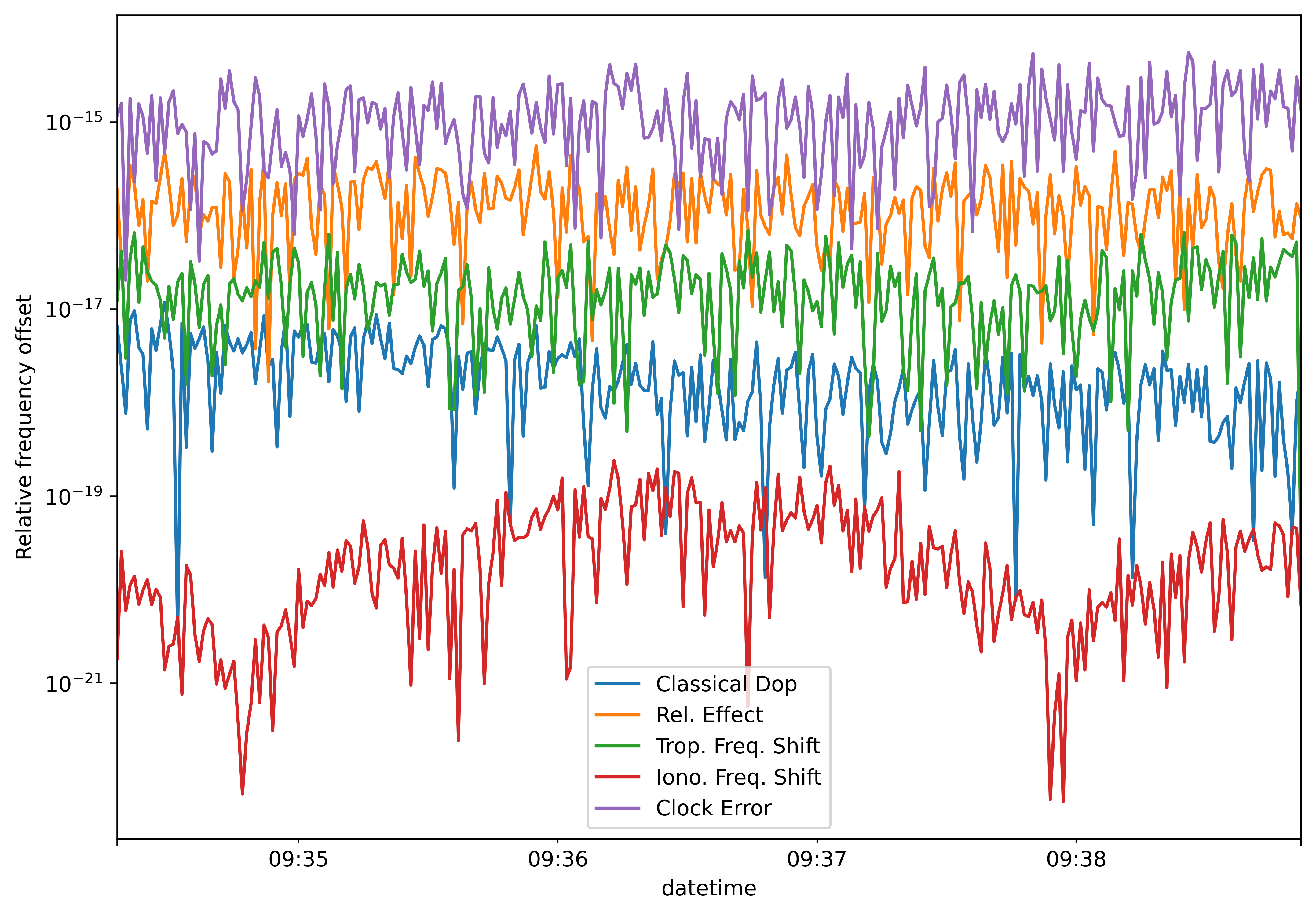}
    \caption{
        Relative frequency offset between theoretical predictions and simulated observations of frequency shift during one slot period (around 5 min, from June 01, 09:34:18 to June 01, 09:38:52). 
        The main error influence factors are denoted in different colors, marked in the left-bottom corner.
       The most prominent error source is the clock error, at least two magnitudes larger than other error influences.}
    \label{fig4}
\end{figure}

There are four sections (periods) in which observation data are available (see the solid black line in Fig.~\ref{fig2}) during the one day's experiment. {The result of one-day experiment {is} {$0.17 \pm 3.05 \times 10^{-16}$}, where {$0.17\times 10^{-16}$} is the mean offset between {the true value and the experiment result, corresponding to} $1.68\times 10^{-7}$ for the offset of $\alpha$, {and} {$\pm 3.05 \times 10^{-16}$} is the {uncertainty (STD).}} The offset of $\alpha$ reflects the precision of the {GRS} test experiment. However, the result of only one simulation experiment might be influenced by accidental factors. Therefore, we repeated the above experiment 40 times for the same day under different random seeds of errors generation to estimate the precision of the {GRS} test, and the results of $\alpha$ offset of 40 experiments are depicted in Fig.~\ref{fig5}.
We can see that the mean offset of $\alpha$ is $2.7\times 10^{-8}$, with the standard deviation of $2.15\times 10^{-7}$.

\begin{figure}[htbp]
    \includegraphics[width=8.6cm,keepaspectratio]{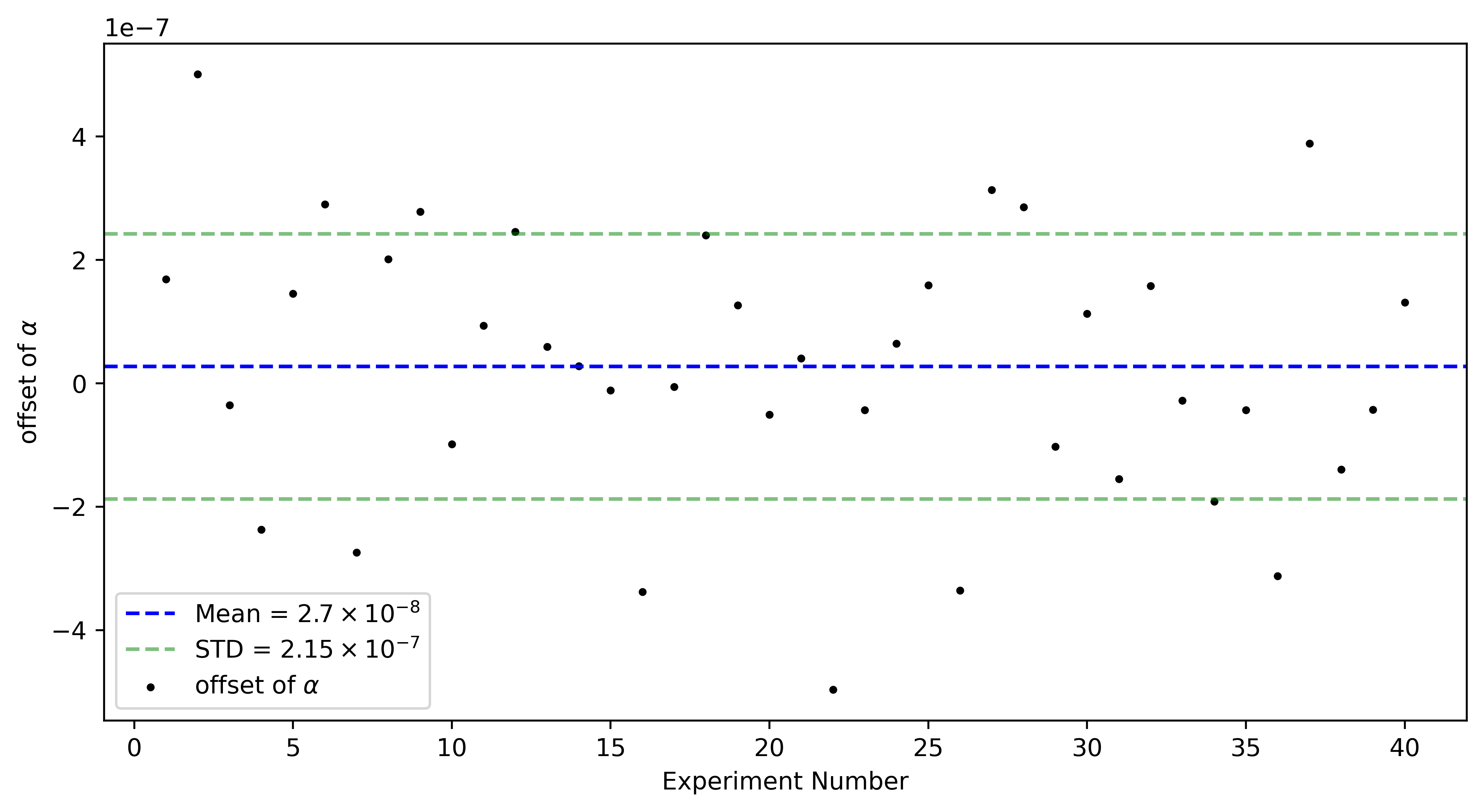}
    \caption{\label{fig5}
        The results of 40 different simulation experiments. In each experiment, we use the same parameters as listed in Table~\ref{tab:table1} with different random seeds of errors generation.
    }
\end{figure}

\section{Conclusion}\label{V}
Based on the uplink and downlink frequency signals in free space between the CSS and a ground station, we established a formulation for determining the gravitational potential difference between the CSS and the ground station.
Based on this formulation, simulation results have shown that the {GRS} could be tested at around $2\times 10^{-7}$, at least two orders in magnitude higher than the presently highest accuracy level. In addition, the proposed formulation in this study could be a new approach applied to determining the geopotential difference between arbitrary two ground stations. 
In contrast, the conventional approach is gravimetry plus leveling~\cite{Heiskanen1967}, which is laborious and low-efficient~\cite{Shen2020}.
Suppose two ground stations equipped with optical clocks with the stability of $1\times 10^{-18}$ can simultaneously observe the frequency signals from CSS, the accuracy of determining the geopotential difference between the two ground stations could achieve a level of $0.1$ m$^2$/s$^2$, because subtraction of the observation between CSS and one station and that between CSS and another station will cancel the common error sources stemming from CSS. With the rapid development of time-frequency science, the optical-atomic clocks with unprecedented accuracy and stability have been generated~\cite{Nicholson2015,Huntemann2016,McGrew2018,Oelker2019,Nakamura2020,Brewer2019}.
It dramatically expands the clocks' application scopes in various branches.
Besides its application in geopotential determination, this formulation can also be applied in height propagation between two datum stations separated by oceans, unifying the vertical height systems around the world~\cite{Shen2020}.

\begin{acknowledgments}
    This study is supported by the National Natural Science Foundation of China (NSFC) (Grant Nos. 42030105, 41721003, 41631072, 41874023, 41804012, 41974034), Space Station Project (Grant No. 2020-228), and Natural Science Foundation of Hubei Province (Grant No. 2019CFB611).
\end{acknowledgments}

\normalem
\bibliographystyle{apsrev4-1}
\bibliography{Shen_ref}

\begin{thebibliography}{39}%
\makeatletter
\providecommand \@ifxundefined [1]{%
 \@ifx{#1\undefined}
}%
\providecommand \@ifnum [1]{%
 \ifnum #1\expandafter \@firstoftwo
 \else \expandafter \@secondoftwo
 \fi
}%
\providecommand \@ifx [1]{%
 \ifx #1\expandafter \@firstoftwo
 \else \expandafter \@secondoftwo
 \fi
}%
\providecommand \natexlab [1]{#1}%
\providecommand \enquote  [1]{``#1''}%
\providecommand \bibnamefont  [1]{#1}%
\providecommand \bibfnamefont [1]{#1}%
\providecommand \citenamefont [1]{#1}%
\providecommand \href@noop [0]{\@secondoftwo}%
\providecommand \href [0]{\begingroup \@sanitize@url \@href}%
\providecommand \@href[1]{\@@startlink{#1}\@@href}%
\providecommand \@@href[1]{\endgroup#1\@@endlink}%
\providecommand \@sanitize@url [0]{\catcode `\\12\catcode `\$12\catcode
  `\&12\catcode `\#12\catcode `\^12\catcode `\_12\catcode `\%12\relax}%
\providecommand \@@startlink[1]{}%
\providecommand \@@endlink[0]{}%
\providecommand \url  [0]{\begingroup\@sanitize@url \@url }%
\providecommand \@url [1]{\endgroup\@href {#1}{\urlprefix }}%
\providecommand \urlprefix  [0]{URL }%
\providecommand \Eprint [0]{\href }%
\providecommand \doibase [0]{http://dx.doi.org/}%
\providecommand \selectlanguage [0]{\@gobble}%
\providecommand \bibinfo  [0]{\@secondoftwo}%
\providecommand \bibfield  [0]{\@secondoftwo}%
\providecommand \translation [1]{[#1]}%
\providecommand \BibitemOpen [0]{}%
\providecommand \bibitemStop [0]{}%
\providecommand \bibitemNoStop [0]{.\EOS\space}%
\providecommand \EOS [0]{\spacefactor3000\relax}%
\providecommand \BibitemShut  [1]{\csname bibitem#1\endcsname}%
\let\auto@bib@innerbib\@empty
\bibitem [{\citenamefont {Vessot}\ and\ \citenamefont
  {Levine}(1979)}]{Vessot1979}%
  \BibitemOpen
  \bibfield  {author} {\bibinfo {author} {\bibfnamefont {R.~F.~C.}\
  \bibnamefont {Vessot}}\ and\ \bibinfo {author} {\bibfnamefont {M.~W.}\
  \bibnamefont {Levine}},\ }\href@noop {} {\bibfield  {journal} {\bibinfo
  {journal} {General relativity and gravitation}\ }\textbf {\bibinfo {volume}
  {10}},\ \bibinfo {pages} {181} (\bibinfo {year} {1979})}\BibitemShut
  {NoStop}%
\bibitem [{\citenamefont {Takamoto}\ \emph {et~al.}(2020)\citenamefont
  {Takamoto}, \citenamefont {Ushijima}, \citenamefont {Ohmae}, \citenamefont
  {Yahagi}, \citenamefont {Kokado}, \citenamefont {Shinkai},\ and\
  \citenamefont {Katori}}]{Takamoto2020-rb}%
  \BibitemOpen
  \bibfield  {author} {\bibinfo {author} {\bibfnamefont {M.}~\bibnamefont
  {Takamoto}}, \bibinfo {author} {\bibfnamefont {I.}~\bibnamefont {Ushijima}},
  \bibinfo {author} {\bibfnamefont {N.}~\bibnamefont {Ohmae}}, \bibinfo
  {author} {\bibfnamefont {T.}~\bibnamefont {Yahagi}}, \bibinfo {author}
  {\bibfnamefont {K.}~\bibnamefont {Kokado}}, \bibinfo {author} {\bibfnamefont
  {H.}~\bibnamefont {Shinkai}}, \ and\ \bibinfo {author} {\bibfnamefont
  {H.}~\bibnamefont {Katori}},\ }\href@noop {} {\bibfield  {journal} {\bibinfo
  {journal} {Nature Photonics}\ }\textbf {\bibinfo {volume} {14}},\ \bibinfo
  {pages} {411} (\bibinfo {year} {2020})}\BibitemShut {NoStop}%
\bibitem [{\citenamefont {Vessot}\ \emph {et~al.}(1980)\citenamefont {Vessot},
  \citenamefont {Levine}, \citenamefont {Mattison}, \citenamefont {Blomberg},
  \citenamefont {Hoffman}, \citenamefont {Nystrom}, \citenamefont {Farrell},
  \citenamefont {Decher}, \citenamefont {Eby}, \citenamefont {Baugher},
  \citenamefont {Watts}, \citenamefont {Teuber},\ and\ \citenamefont
  {Wills}}]{Vessot1980}%
  \BibitemOpen
  \bibfield  {author} {\bibinfo {author} {\bibfnamefont {R.~F.~C.}\
  \bibnamefont {Vessot}}, \bibinfo {author} {\bibfnamefont {M.~W.}\
  \bibnamefont {Levine}}, \bibinfo {author} {\bibfnamefont {E.~M.}\
  \bibnamefont {Mattison}}, \bibinfo {author} {\bibfnamefont {E.~L.}\
  \bibnamefont {Blomberg}}, \bibinfo {author} {\bibfnamefont {T.~E.}\
  \bibnamefont {Hoffman}}, \bibinfo {author} {\bibfnamefont {G.~U.}\
  \bibnamefont {Nystrom}}, \bibinfo {author} {\bibfnamefont {B.~F.}\
  \bibnamefont {Farrell}}, \bibinfo {author} {\bibfnamefont {R.}~\bibnamefont
  {Decher}}, \bibinfo {author} {\bibfnamefont {P.~B.}\ \bibnamefont {Eby}},
  \bibinfo {author} {\bibfnamefont {C.~R.}\ \bibnamefont {Baugher}}, \bibinfo
  {author} {\bibfnamefont {J.~W.}\ \bibnamefont {Watts}}, \bibinfo {author}
  {\bibfnamefont {D.~L.}\ \bibnamefont {Teuber}}, \ and\ \bibinfo {author}
  {\bibfnamefont {F.~D.}\ \bibnamefont {Wills}},\ }\href {\doibase
  10.1103/PhysRevLett.45.2081} {\bibfield  {journal} {\bibinfo  {journal}
  {Physical Review Letters}\ }\textbf {\bibinfo {volume} {45}},\ \bibinfo
  {pages} {2081} (\bibinfo {year} {1980})}\BibitemShut {NoStop}%
\bibitem [{\citenamefont {Kouba}(2021)}]{Kouba2021}%
  \BibitemOpen
  \bibfield  {author} {\bibinfo {author} {\bibfnamefont {J.}~\bibnamefont
  {Kouba}},\ }\href@noop {} {\bibfield  {journal} {\bibinfo  {journal} {GPS
  Solutions}\ }\textbf {\bibinfo {volume} {25}},\ \bibinfo {pages} {1}
  (\bibinfo {year} {2021})}\BibitemShut {NoStop}%
\bibitem [{\citenamefont {Delva}\ \emph {et~al.}(2018)\citenamefont {Delva},
  \citenamefont {Puchades}, \citenamefont {Sch\"onemann}, \citenamefont
  {Dilssner}, \citenamefont {Courde}, \citenamefont {Bertone}, \citenamefont
  {Gonzalez}, \citenamefont {Hees}, \citenamefont {Le~Poncin-Lafitte},
  \citenamefont {Meynadier}, \citenamefont {Prieto-Cerdeira}, \citenamefont
  {Sohet}, \citenamefont {Ventura-Traveset},\ and\ \citenamefont
  {Wolf}}]{Delva2018}%
  \BibitemOpen
  \bibfield  {author} {\bibinfo {author} {\bibfnamefont {P.}~\bibnamefont
  {Delva}}, \bibinfo {author} {\bibfnamefont {N.}~\bibnamefont {Puchades}},
  \bibinfo {author} {\bibfnamefont {E.}~\bibnamefont {Sch\"onemann}}, \bibinfo
  {author} {\bibfnamefont {F.}~\bibnamefont {Dilssner}}, \bibinfo {author}
  {\bibfnamefont {C.}~\bibnamefont {Courde}}, \bibinfo {author} {\bibfnamefont
  {S.}~\bibnamefont {Bertone}}, \bibinfo {author} {\bibfnamefont
  {F.}~\bibnamefont {Gonzalez}}, \bibinfo {author} {\bibfnamefont
  {A.}~\bibnamefont {Hees}}, \bibinfo {author} {\bibfnamefont {C.}~\bibnamefont
  {Le~Poncin-Lafitte}}, \bibinfo {author} {\bibfnamefont {F.}~\bibnamefont
  {Meynadier}}, \bibinfo {author} {\bibfnamefont {R.}~\bibnamefont
  {Prieto-Cerdeira}}, \bibinfo {author} {\bibfnamefont {B.}~\bibnamefont
  {Sohet}}, \bibinfo {author} {\bibfnamefont {J.}~\bibnamefont
  {Ventura-Traveset}}, \ and\ \bibinfo {author} {\bibfnamefont
  {P.}~\bibnamefont {Wolf}},\ }\href {\doibase 10.1103/PhysRevLett.121.231101}
  {\bibfield  {journal} {\bibinfo  {journal} {Phys. Rev. Lett.}\ }\textbf
  {\bibinfo {volume} {121}},\ \bibinfo {pages} {231101} (\bibinfo {year}
  {2018})}\BibitemShut {NoStop}%
\bibitem [{\citenamefont {Herrmann}\ \emph {et~al.}(2018)\citenamefont
  {Herrmann}, \citenamefont {Finke}, \citenamefont {L{\"u}lf}, \citenamefont
  {Kichakova}, \citenamefont {Puetzfeld}, \citenamefont {Knickmann},
  \citenamefont {List}, \citenamefont {Rievers}, \citenamefont {Giorgi},
  \citenamefont {G{\"u}nther}, \citenamefont {Dittus}, \citenamefont
  {Prieto-Cerdeira}, \citenamefont {Dilssner}, \citenamefont {Gonzalez},
  \citenamefont {Schonemann}, \citenamefont {Ventura-Traveset},\ and\
  \citenamefont {Lammerzahl}}]{Herrmann2018}%
  \BibitemOpen
  \bibfield  {author} {\bibinfo {author} {\bibfnamefont {S.}~\bibnamefont
  {Herrmann}}, \bibinfo {author} {\bibfnamefont {F.}~\bibnamefont {Finke}},
  \bibinfo {author} {\bibfnamefont {M.}~\bibnamefont {L{\"u}lf}}, \bibinfo
  {author} {\bibfnamefont {O.}~\bibnamefont {Kichakova}}, \bibinfo {author}
  {\bibfnamefont {D.}~\bibnamefont {Puetzfeld}}, \bibinfo {author}
  {\bibfnamefont {D.}~\bibnamefont {Knickmann}}, \bibinfo {author}
  {\bibfnamefont {M.}~\bibnamefont {List}}, \bibinfo {author} {\bibfnamefont
  {B.}~\bibnamefont {Rievers}}, \bibinfo {author} {\bibfnamefont
  {G.}~\bibnamefont {Giorgi}}, \bibinfo {author} {\bibfnamefont
  {C.}~\bibnamefont {G{\"u}nther}}, \bibinfo {author} {\bibfnamefont
  {H.}~\bibnamefont {Dittus}}, \bibinfo {author} {\bibfnamefont
  {R.}~\bibnamefont {Prieto-Cerdeira}}, \bibinfo {author} {\bibfnamefont
  {F.}~\bibnamefont {Dilssner}}, \bibinfo {author} {\bibfnamefont
  {F.}~\bibnamefont {Gonzalez}}, \bibinfo {author} {\bibfnamefont
  {E.}~\bibnamefont {Schonemann}}, \bibinfo {author} {\bibfnamefont
  {J.}~\bibnamefont {Ventura-Traveset}}, \ and\ \bibinfo {author}
  {\bibfnamefont {C.}~\bibnamefont {Lammerzahl}},\ }\href {\doibase
  10.1103/PhysRevLett.121.231102} {\bibfield  {journal} {\bibinfo  {journal}
  {Physical review letters}\ }\textbf {\bibinfo {volume} {121}},\ \bibinfo
  {pages} {231102} (\bibinfo {year} {2018})}\BibitemShut {NoStop}%
\bibitem [{\citenamefont {Cacciapuoti}\ and\ \citenamefont
  {Salomon}(2011)}]{cacciapuoti2011}%
  \BibitemOpen
  \bibfield  {author} {\bibinfo {author} {\bibfnamefont {L.}~\bibnamefont
  {Cacciapuoti}}\ and\ \bibinfo {author} {\bibfnamefont {C.}~\bibnamefont
  {Salomon}},\ }in\ \href@noop {} {\emph {\bibinfo {booktitle} {Journal of
  Physics: Conference Series}}},\ Vol.\ \bibinfo {volume} {327}\ (\bibinfo
  {organization} {IOP Publishing},\ \bibinfo {year} {2011})\ p.\ \bibinfo
  {pages} {012049}\BibitemShut {NoStop}%
\bibitem [{\citenamefont {Meynadier}\ \emph {et~al.}(2018)\citenamefont
  {Meynadier}, \citenamefont {Delva}, \citenamefont {le~Poncin-Lafitte},
  \citenamefont {Guerlin},\ and\ \citenamefont {Wolf}}]{meynadier2018}%
  \BibitemOpen
  \bibfield  {author} {\bibinfo {author} {\bibfnamefont {F.}~\bibnamefont
  {Meynadier}}, \bibinfo {author} {\bibfnamefont {P.}~\bibnamefont {Delva}},
  \bibinfo {author} {\bibfnamefont {C.}~\bibnamefont {le~Poncin-Lafitte}},
  \bibinfo {author} {\bibfnamefont {C.}~\bibnamefont {Guerlin}}, \ and\
  \bibinfo {author} {\bibfnamefont {P.}~\bibnamefont {Wolf}},\ }\href@noop {}
  {\bibfield  {journal} {\bibinfo  {journal} {Classical and Quantum Gravity}\
  }\textbf {\bibinfo {volume} {35}},\ \bibinfo {pages} {035018} (\bibinfo
  {year} {2018})}\BibitemShut {NoStop}%
\bibitem [{\citenamefont {Blanchet}\ \emph {et~al.}(2001)\citenamefont
  {Blanchet}, \citenamefont {Salomon}, \citenamefont {Teyssandier},\ and\
  \citenamefont {Wolf}}]{blanchet2001}%
  \BibitemOpen
  \bibfield  {author} {\bibinfo {author} {\bibfnamefont {L.}~\bibnamefont
  {Blanchet}}, \bibinfo {author} {\bibfnamefont {C.}~\bibnamefont {Salomon}},
  \bibinfo {author} {\bibfnamefont {P.}~\bibnamefont {Teyssandier}}, \ and\
  \bibinfo {author} {\bibfnamefont {P.}~\bibnamefont {Wolf}},\ }\href@noop {}
  {\bibfield  {journal} {\bibinfo  {journal} {Astronomy \& Astrophysics}\
  }\textbf {\bibinfo {volume} {370}},\ \bibinfo {pages} {320} (\bibinfo {year}
  {2001})}\BibitemShut {NoStop}%
\bibitem [{\citenamefont {Linet}\ and\ \citenamefont
  {Teyssandier}(2002)}]{linet2002}%
  \BibitemOpen
  \bibfield  {author} {\bibinfo {author} {\bibfnamefont {B.}~\bibnamefont
  {Linet}}\ and\ \bibinfo {author} {\bibfnamefont {P.}~\bibnamefont
  {Teyssandier}},\ }\href {\doibase 10.1103/PhysRevD.66.024045} {\bibfield
  {journal} {\bibinfo  {journal} {Phys. Rev. D}\ }\textbf {\bibinfo {volume}
  {66}},\ \bibinfo {pages} {024045} (\bibinfo {year} {2002})}\BibitemShut
  {NoStop}%
\bibitem [{\citenamefont {Sun}\ \emph {et~al.}(2021)\citenamefont {Sun},
  \citenamefont {Shen}, \citenamefont {Shen}, \citenamefont {Cai},
  \citenamefont {Xu},\ and\ \citenamefont {Zhang}}]{Sun2021}%
  \BibitemOpen
  \bibfield  {author} {\bibinfo {author} {\bibfnamefont {X.}~\bibnamefont
  {Sun}}, \bibinfo {author} {\bibfnamefont {W.-B.}\ \bibnamefont {Shen}},
  \bibinfo {author} {\bibfnamefont {Z.}~\bibnamefont {Shen}}, \bibinfo {author}
  {\bibfnamefont {C.}~\bibnamefont {Cai}}, \bibinfo {author} {\bibfnamefont
  {W.}~\bibnamefont {Xu}}, \ and\ \bibinfo {author} {\bibfnamefont
  {P.}~\bibnamefont {Zhang}},\ }\href {\doibase
  10.1140/epjc/s10052-021-09415-y} {\bibfield  {journal} {\bibinfo  {journal}
  {The European Physical Journal C}\ }\textbf {\bibinfo {volume} {81}},\
  \bibinfo {pages} {1} (\bibinfo {year} {2021})}\BibitemShut {NoStop}%
\bibitem [{\citenamefont {Shen}\ \emph {et~al.}(2021)\citenamefont {Shen},
  \citenamefont {Shen}, \citenamefont {Zhang}, \citenamefont {He},
  \citenamefont {Cai}, \citenamefont {Tian},\ and\ \citenamefont
  {Zhang}}]{Shen2021}%
  \BibitemOpen
  \bibfield  {author} {\bibinfo {author} {\bibfnamefont {Z.}~\bibnamefont
  {Shen}}, \bibinfo {author} {\bibfnamefont {W.-B.}\ \bibnamefont {Shen}},
  \bibinfo {author} {\bibfnamefont {T.}~\bibnamefont {Zhang}}, \bibinfo
  {author} {\bibfnamefont {L.}~\bibnamefont {He}}, \bibinfo {author}
  {\bibfnamefont {Z.}~\bibnamefont {Cai}}, \bibinfo {author} {\bibfnamefont
  {X.}~\bibnamefont {Tian}}, \ and\ \bibinfo {author} {\bibfnamefont
  {P.}~\bibnamefont {Zhang}},\ }\href {\doibase 10.1016/j.asr.2021.07.004}
  {\bibfield  {journal} {\bibinfo  {journal} {Advances in Space Research}\
  }\textbf {\bibinfo {volume} {68}},\ \bibinfo {pages} {2776} (\bibinfo {year}
  {2021})}\BibitemShut {NoStop}%
\bibitem [{\citenamefont {Guo}(2021)}]{Guo2021}%
  \BibitemOpen
  \bibfield  {author} {\bibinfo {author} {\bibfnamefont {Y.~M.}\ \bibnamefont
  {Guo}},\ }in\ \href@noop {} {\emph {\bibinfo {booktitle} {The 12th China
  Satellite Navigation Conference, CSNC 2021}}}\ (\bibinfo {year}
  {2021})\BibitemShut {NoStop}%
\bibitem [{\citenamefont {Wang}(2021)}]{Wang2021}%
  \BibitemOpen
  \bibfield  {author} {\bibinfo {author} {\bibfnamefont {W.~B.}\ \bibnamefont
  {Wang}},\ }in\ \href@noop {} {\emph {\bibinfo {booktitle} {The 12th China
  Satellite Navigation Conference, CSNC 2021}}}\ (\bibinfo {year}
  {2021})\BibitemShut {NoStop}%
\bibitem [{\citenamefont {Elliot}(1983)}]{IEEE1983}%
  \BibitemOpen
  \bibfield  {author} {\bibinfo {author} {\bibfnamefont {R.~S.}\ \bibnamefont
  {Elliot}},\ }\href@noop {} {\bibfield  {journal} {\bibinfo  {journal} {IEEE
  Transactions on Antennas and Propagation}\ }\textbf {\bibinfo {volume}
  {31}},\ \bibinfo {pages} {1} (\bibinfo {year} {1983})}\BibitemShut {NoStop}%
\bibitem [{\citenamefont {Hoque}\ and\ \citenamefont
  {Jakowski}(2012)}]{Hoque2012}%
  \BibitemOpen
  \bibfield  {author} {\bibinfo {author} {\bibfnamefont {M.~M.}\ \bibnamefont
  {Hoque}}\ and\ \bibinfo {author} {\bibfnamefont {N.}~\bibnamefont
  {Jakowski}},\ }\href {\doibase 10.5772/30090} {\bibfield  {journal} {\bibinfo
   {journal} {Global navigation satellite systems: signal, theory and
  applications}\ }\textbf {\bibinfo {volume} {10}},\ \bibinfo {pages} {30090}
  (\bibinfo {year} {2012})}\BibitemShut {NoStop}%
\bibitem [{\citenamefont {Sch{\"u}ller}(2020)}]{Schuller2020}%
  \BibitemOpen
  \bibfield  {author} {\bibinfo {author} {\bibfnamefont {K.}~\bibnamefont
  {Sch{\"u}ller}},\ }\href {http://ggp.bkg.bund.de/eterna?download=7283} {\emph
  {\bibinfo {title} {Program System ETERNA-x et34-x-v80-* for Earth and Ocean
  Tides Analysis and Prediction}}} (\bibinfo {year} {2020})\BibitemShut
  {NoStop}%
\bibitem [{\citenamefont {Hartmann}\ and\ \citenamefont
  {Wenzel}(1995)}]{Hartmann1995}%
  \BibitemOpen
  \bibfield  {author} {\bibinfo {author} {\bibfnamefont {T.}~\bibnamefont
  {Hartmann}}\ and\ \bibinfo {author} {\bibfnamefont {H.-G.}\ \bibnamefont
  {Wenzel}},\ }\href@noop {} {\bibfield  {journal} {\bibinfo  {journal}
  {Geophysical research letters}\ }\textbf {\bibinfo {volume} {22}},\ \bibinfo
  {pages} {3553} (\bibinfo {year} {1995})}\BibitemShut {NoStop}%
\bibitem [{\citenamefont {Wenzel}(1997)}]{Wenzel1997}%
  \BibitemOpen
  \bibfield  {author} {\bibinfo {author} {\bibfnamefont {H.-G.}\ \bibnamefont
  {Wenzel}},\ }\href {https://doi.org/10.1007/BFb0011455} {\bibfield  {journal}
  {\bibinfo  {journal} {Tidal phenomena}\ ,\ \bibinfo {pages} {9}} (\bibinfo
  {year} {1997})}\BibitemShut {NoStop}%
\bibitem [{\citenamefont {Hoffmann}(1961)}]{Hoffmann1961}%
  \BibitemOpen
  \bibfield  {author} {\bibinfo {author} {\bibfnamefont {B.}~\bibnamefont
  {Hoffmann}},\ }\href {\doibase 10.1103/PhysRev.121.337} {\bibfield  {journal}
  {\bibinfo  {journal} {Physical Review}\ }\textbf {\bibinfo {volume} {121}},\
  \bibinfo {pages} {337} (\bibinfo {year} {1961})}\BibitemShut {NoStop}%
\bibitem [{\citenamefont {Kleppner}\ \emph {et~al.}(1970)\citenamefont
  {Kleppner}, \citenamefont {Vessot},\ and\ \citenamefont
  {Ramsey}}]{Kleppner1970-db}%
  \BibitemOpen
  \bibfield  {author} {\bibinfo {author} {\bibfnamefont {D.}~\bibnamefont
  {Kleppner}}, \bibinfo {author} {\bibfnamefont {R.~F.~C.}\ \bibnamefont
  {Vessot}}, \ and\ \bibinfo {author} {\bibfnamefont {N.~F.}\ \bibnamefont
  {Ramsey}},\ }\href@noop {} {\bibfield  {journal} {\bibinfo  {journal}
  {Astrophysics and Space Science}\ }\textbf {\bibinfo {volume} {6}},\ \bibinfo
  {pages} {13} (\bibinfo {year} {1970})}\BibitemShut {NoStop}%
\bibitem [{\citenamefont {Saastamoinen}(1972)}]{Saastamoinen1972-de}%
  \BibitemOpen
  \bibfield  {author} {\bibinfo {author} {\bibfnamefont {J.}~\bibnamefont
  {Saastamoinen}},\ }\href@noop {} {\bibfield  {journal} {\bibinfo  {journal}
  {The use of artificial satellites for geodesy}\ }\textbf {\bibinfo {volume}
  {15}},\ \bibinfo {pages} {247} (\bibinfo {year} {1972})}\BibitemShut
  {NoStop}%
\bibitem [{\citenamefont {B{\"o}hm}\ and\ \citenamefont
  {Schuh}(2003)}]{Boehm2013}%
  \BibitemOpen
  \bibfield  {author} {\bibinfo {author} {\bibfnamefont {J.}~\bibnamefont
  {B{\"o}hm}}\ and\ \bibinfo {author} {\bibfnamefont {H.}~\bibnamefont
  {Schuh}},\ }\href {http://www.evga.org/files/2003EVGA-proc_Leipzig.pdf}
  {\emph {\bibinfo {title} {Vienna mapping functions}}}\ (\bibinfo  {publisher}
  {na},\ \bibinfo {year} {2003})\BibitemShut {NoStop}%
\bibitem [{\citenamefont {Bilitza}\ \emph {et~al.}(1993)\citenamefont
  {Bilitza}, \citenamefont {Rawer}, \citenamefont {Bossy},\ and\ \citenamefont
  {Gulyaeva}}]{Bilitza1993}%
  \BibitemOpen
  \bibfield  {author} {\bibinfo {author} {\bibfnamefont {D.}~\bibnamefont
  {Bilitza}}, \bibinfo {author} {\bibfnamefont {K.}~\bibnamefont {Rawer}},
  \bibinfo {author} {\bibfnamefont {L.}~\bibnamefont {Bossy}}, \ and\ \bibinfo
  {author} {\bibfnamefont {T.}~\bibnamefont {Gulyaeva}},\ }\href {\doibase
  10.1016/0273-1177(93)90241-3} {\bibfield  {journal} {\bibinfo  {journal}
  {Advances in Space Research}\ }\textbf {\bibinfo {volume} {13}},\ \bibinfo
  {pages} {15} (\bibinfo {year} {1993})}\BibitemShut {NoStop}%
\bibitem [{\citenamefont {Baerenzung}\ \emph {et~al.}(2020)\citenamefont
  {Baerenzung}, \citenamefont {Holschneider}, \citenamefont {Wicht},
  \citenamefont {Lesur},\ and\ \citenamefont {Sanchez}}]{Baerenzung2020}%
  \BibitemOpen
  \bibfield  {author} {\bibinfo {author} {\bibfnamefont {J.}~\bibnamefont
  {Baerenzung}}, \bibinfo {author} {\bibfnamefont {M.}~\bibnamefont
  {Holschneider}}, \bibinfo {author} {\bibfnamefont {J.}~\bibnamefont {Wicht}},
  \bibinfo {author} {\bibfnamefont {V.}~\bibnamefont {Lesur}}, \ and\ \bibinfo
  {author} {\bibfnamefont {S.}~\bibnamefont {Sanchez}},\ }\href {\doibase
  10.1186/s40623-020-01295-y} {\bibfield  {journal} {\bibinfo  {journal}
  {Earth, Planets and Space}\ }\textbf {\bibinfo {volume} {72}},\ \bibinfo
  {pages} {1} (\bibinfo {year} {2020})}\BibitemShut {NoStop}%
\bibitem [{\citenamefont {Petit}\ and\ \citenamefont
  {Luzum}(2010)}]{Petit2010}%
  \BibitemOpen
  \bibfield  {author} {\bibinfo {author} {\bibfnamefont {G.}~\bibnamefont
  {Petit}}\ and\ \bibinfo {author} {\bibfnamefont {B.}~\bibnamefont {Luzum}},\
  }\href@noop {} {\emph {\bibinfo {title} {IERS conventions (2010)}}},\
  \bibinfo {type} {Report}\ (\bibinfo  {institution} {Bureau International des
  Poids et mesures sevres (france)},\ \bibinfo {year} {2010})\BibitemShut
  {NoStop}%
\bibitem [{\citenamefont {Pavlis}\ \emph {et~al.}(2008)\citenamefont {Pavlis},
  \citenamefont {Holmes}, \citenamefont {Kenyon},\ and\ \citenamefont
  {Factor}}]{Pavlis2008}%
  \BibitemOpen
  \bibfield  {author} {\bibinfo {author} {\bibfnamefont {N.~K.}\ \bibnamefont
  {Pavlis}}, \bibinfo {author} {\bibfnamefont {S.~A.}\ \bibnamefont {Holmes}},
  \bibinfo {author} {\bibfnamefont {S.~C.}\ \bibnamefont {Kenyon}}, \ and\
  \bibinfo {author} {\bibfnamefont {J.~K.}\ \bibnamefont {Factor}},\ }in\
  \href@noop {} {\emph {\bibinfo {booktitle} {AGU Fall Meeting Abstracts}}}\
  (\bibinfo {year} {2008})\BibitemShut {NoStop}%
\bibitem [{\citenamefont {Weiping}\ \emph {et~al.}(2012)\citenamefont
  {Weiping}, \citenamefont {Hongxing},\ and\ \citenamefont {Qi}}]{Li2012}%
  \BibitemOpen
  \bibfield  {author} {\bibinfo {author} {\bibfnamefont {L.}~\bibnamefont
  {Weiping}}, \bibinfo {author} {\bibfnamefont {Q.}~\bibnamefont {Hongxing}}, \
  and\ \bibinfo {author} {\bibfnamefont {D.}~\bibnamefont {Qi}},\ }in\
  \href@noop {} {\emph {\bibinfo {booktitle} {2012 IEEE International
  Conference on Information and Automation}}}\ (\bibinfo {organization}
  {IEEE},\ \bibinfo {year} {2012})\ pp.\ \bibinfo {pages}
  {811--813}\BibitemShut {NoStop}%
\bibitem [{\citenamefont {Pavlis}\ \emph {et~al.}(2012)\citenamefont {Pavlis},
  \citenamefont {Holmes}, \citenamefont {Kenyon},\ and\ \citenamefont
  {Factor}}]{Pavlis2012}%
  \BibitemOpen
  \bibfield  {author} {\bibinfo {author} {\bibfnamefont {N.~K.}\ \bibnamefont
  {Pavlis}}, \bibinfo {author} {\bibfnamefont {S.~A.}\ \bibnamefont {Holmes}},
  \bibinfo {author} {\bibfnamefont {S.~C.}\ \bibnamefont {Kenyon}}, \ and\
  \bibinfo {author} {\bibfnamefont {J.~K.}\ \bibnamefont {Factor}},\ }\href
  {\doibase 10.1029/2011JB008916} {\bibfield  {journal} {\bibinfo  {journal}
  {Journal of geophysical research: solid earth}\ }\textbf {\bibinfo {volume}
  {117}},\ \bibinfo {pages} {1} (\bibinfo {year} {2012})}\BibitemShut {NoStop}%
\bibitem [{\citenamefont {Allan}\ \emph {et~al.}(1991)\citenamefont {Allan},
  \citenamefont {Weiss},\ and\ \citenamefont {Jespersen}}]{Allan1991}%
  \BibitemOpen
  \bibfield  {author} {\bibinfo {author} {\bibfnamefont {D.~W.}\ \bibnamefont
  {Allan}}, \bibinfo {author} {\bibfnamefont {M.~A.}\ \bibnamefont {Weiss}}, \
  and\ \bibinfo {author} {\bibfnamefont {J.~L.}\ \bibnamefont {Jespersen}},\
  }in\ \href {\doibase 10.1109/freq.1991.145966} {\emph {\bibinfo {booktitle}
  {Proceedings of the 45th Annual Symposium on Frequency Control 1991}}}\
  (\bibinfo {organization} {IEEE},\ \bibinfo {year} {1991})\ pp.\ \bibinfo
  {pages} {667--678}\BibitemShut {NoStop}%
\bibitem [{\citenamefont {Bilitza}\ \emph {et~al.}(2017)\citenamefont
  {Bilitza}, \citenamefont {Altadill}, \citenamefont {Truhlik}, \citenamefont
  {Shubin}, \citenamefont {Galkin}, \citenamefont {Reinisch},\ and\
  \citenamefont {Huang}}]{Bilitza2017}%
  \BibitemOpen
  \bibfield  {author} {\bibinfo {author} {\bibfnamefont {D.}~\bibnamefont
  {Bilitza}}, \bibinfo {author} {\bibfnamefont {D.}~\bibnamefont {Altadill}},
  \bibinfo {author} {\bibfnamefont {V.}~\bibnamefont {Truhlik}}, \bibinfo
  {author} {\bibfnamefont {V.}~\bibnamefont {Shubin}}, \bibinfo {author}
  {\bibfnamefont {I.}~\bibnamefont {Galkin}}, \bibinfo {author} {\bibfnamefont
  {B.}~\bibnamefont {Reinisch}}, \ and\ \bibinfo {author} {\bibfnamefont
  {X.}~\bibnamefont {Huang}},\ }\href {\doibase 10.1002/2016SW001593}
  {\bibfield  {journal} {\bibinfo  {journal} {Space weather}\ }\textbf
  {\bibinfo {volume} {15}},\ \bibinfo {pages} {418} (\bibinfo {year}
  {2017})}\BibitemShut {NoStop}%
\bibitem [{\citenamefont {Heiskanen}\ and\ \citenamefont
  {Moritz}(1967)}]{Heiskanen1967}%
  \BibitemOpen
  \bibfield  {author} {\bibinfo {author} {\bibfnamefont {W.~A.}\ \bibnamefont
  {Heiskanen}}\ and\ \bibinfo {author} {\bibfnamefont {H.}~\bibnamefont
  {Moritz}},\ }\href {\doibase 10.1007/BF02525647} {\emph {\bibinfo {title}
  {Physical geodesy(Book on physical geodesy covering potential theory, gravity
  fields, gravimetric and astrogeodetic methods, statistical analysis, etc)}}}\
  (\bibinfo  {publisher} {W. H. Freeman and Company San Francisco},\ \bibinfo
  {year} {1967})\BibitemShut {NoStop}%
\bibitem [{\citenamefont {Shen}\ \emph {et~al.}(2020)\citenamefont {Shen},
  \citenamefont {Shen},\ and\ \citenamefont {Zhang}}]{Shen2020}%
  \BibitemOpen
  \bibfield  {author} {\bibinfo {author} {\bibfnamefont {Z.}~\bibnamefont
  {Shen}}, \bibinfo {author} {\bibfnamefont {W.-B.}\ \bibnamefont {Shen}}, \
  and\ \bibinfo {author} {\bibfnamefont {S.}~\bibnamefont {Zhang}},\
  }\href@noop {} {\bibfield  {journal} {\bibinfo  {journal} {arXiv preprint
  arXiv:2008.05868}\ ,\ \bibinfo {pages} {1}} (\bibinfo {year}
  {2020})}\BibitemShut {NoStop}%
\bibitem [{\citenamefont {Nicholson}\ \emph {et~al.}(2015)\citenamefont
  {Nicholson}, \citenamefont {Campbell}, \citenamefont {Hutson}, \citenamefont
  {Marti}, \citenamefont {Bloom}, \citenamefont {McNally}, \citenamefont
  {Zhang}, \citenamefont {Barrett}, \citenamefont {Safronova}, \citenamefont
  {Strouse} \emph {et~al.}}]{Nicholson2015}%
  \BibitemOpen
  \bibfield  {author} {\bibinfo {author} {\bibfnamefont {T.~L.}\ \bibnamefont
  {Nicholson}}, \bibinfo {author} {\bibfnamefont {S.~L.}\ \bibnamefont
  {Campbell}}, \bibinfo {author} {\bibfnamefont {R.~B.}\ \bibnamefont
  {Hutson}}, \bibinfo {author} {\bibfnamefont {G.~E.}\ \bibnamefont {Marti}},
  \bibinfo {author} {\bibfnamefont {B.~J.}\ \bibnamefont {Bloom}}, \bibinfo
  {author} {\bibfnamefont {R.~L.}\ \bibnamefont {McNally}}, \bibinfo {author}
  {\bibfnamefont {W.}~\bibnamefont {Zhang}}, \bibinfo {author} {\bibfnamefont
  {M.~D.}\ \bibnamefont {Barrett}}, \bibinfo {author} {\bibfnamefont {M.~S.}\
  \bibnamefont {Safronova}}, \bibinfo {author} {\bibfnamefont {G.~F.}\
  \bibnamefont {Strouse}},  \emph {et~al.},\ }\href {\doibase
  10.1038/ncomms7896} {\bibfield  {journal} {\bibinfo  {journal} {Nature
  communications}\ }\textbf {\bibinfo {volume} {6}},\ \bibinfo {pages} {1}
  (\bibinfo {year} {2015})}\BibitemShut {NoStop}%
\bibitem [{\citenamefont {Huntemann}\ \emph {et~al.}(2016)\citenamefont
  {Huntemann}, \citenamefont {Sanner}, \citenamefont {Lipphardt}, \citenamefont
  {Tamm},\ and\ \citenamefont {Peik}}]{Huntemann2016}%
  \BibitemOpen
  \bibfield  {author} {\bibinfo {author} {\bibfnamefont {N.}~\bibnamefont
  {Huntemann}}, \bibinfo {author} {\bibfnamefont {C.}~\bibnamefont {Sanner}},
  \bibinfo {author} {\bibfnamefont {B.}~\bibnamefont {Lipphardt}}, \bibinfo
  {author} {\bibfnamefont {C.}~\bibnamefont {Tamm}}, \ and\ \bibinfo {author}
  {\bibfnamefont {E.}~\bibnamefont {Peik}},\ }\href {\doibase
  10.1103/PhysRevLett.116.063001} {\bibfield  {journal} {\bibinfo  {journal}
  {Physical review letters}\ }\textbf {\bibinfo {volume} {116}},\ \bibinfo
  {pages} {063001} (\bibinfo {year} {2016})}\BibitemShut {NoStop}%
\bibitem [{\citenamefont {McGrew}\ \emph {et~al.}(2018)\citenamefont {McGrew},
  \citenamefont {Zhang}, \citenamefont {Fasano}, \citenamefont {Sch{\"a}ffer},
  \citenamefont {Beloy}, \citenamefont {Nicolodi}, \citenamefont {Brown},
  \citenamefont {Hinkley}, \citenamefont {Milani}, \citenamefont {Schioppo},
  \citenamefont {Yoon},\ and\ \citenamefont {Ludlow}}]{McGrew2018}%
  \BibitemOpen
  \bibfield  {author} {\bibinfo {author} {\bibfnamefont {W.~F.}\ \bibnamefont
  {McGrew}}, \bibinfo {author} {\bibfnamefont {X.}~\bibnamefont {Zhang}},
  \bibinfo {author} {\bibfnamefont {R.~J.}\ \bibnamefont {Fasano}}, \bibinfo
  {author} {\bibfnamefont {S.~A.}\ \bibnamefont {Sch{\"a}ffer}}, \bibinfo
  {author} {\bibfnamefont {K.}~\bibnamefont {Beloy}}, \bibinfo {author}
  {\bibfnamefont {D.}~\bibnamefont {Nicolodi}}, \bibinfo {author}
  {\bibfnamefont {R.~C.}\ \bibnamefont {Brown}}, \bibinfo {author}
  {\bibfnamefont {N.}~\bibnamefont {Hinkley}}, \bibinfo {author} {\bibfnamefont
  {G.}~\bibnamefont {Milani}}, \bibinfo {author} {\bibfnamefont
  {M.}~\bibnamefont {Schioppo}}, \bibinfo {author} {\bibfnamefont {T.~H.}\
  \bibnamefont {Yoon}}, \ and\ \bibinfo {author} {\bibfnamefont {A.~D.}\
  \bibnamefont {Ludlow}},\ }\href {\doibase 10.1038/s41586-018-0738-2}
  {\bibfield  {journal} {\bibinfo  {journal} {Nature}\ }\textbf {\bibinfo
  {volume} {564}},\ \bibinfo {pages} {87} (\bibinfo {year} {2018})}\BibitemShut
  {NoStop}%
\bibitem [{\citenamefont {Oelker}\ \emph {et~al.}(2019)\citenamefont {Oelker},
  \citenamefont {Hutson}, \citenamefont {Kennedy}, \citenamefont {Sonderhouse},
  \citenamefont {Bothwell}, \citenamefont {Goban}, \citenamefont {Kedar},
  \citenamefont {Sanner}, \citenamefont {Robinson}, \citenamefont {Marti} \emph
  {et~al.}}]{Oelker2019}%
  \BibitemOpen
  \bibfield  {author} {\bibinfo {author} {\bibfnamefont {E.}~\bibnamefont
  {Oelker}}, \bibinfo {author} {\bibfnamefont {R.~B.}\ \bibnamefont {Hutson}},
  \bibinfo {author} {\bibfnamefont {C.~J.}\ \bibnamefont {Kennedy}}, \bibinfo
  {author} {\bibfnamefont {L.}~\bibnamefont {Sonderhouse}}, \bibinfo {author}
  {\bibfnamefont {T.}~\bibnamefont {Bothwell}}, \bibinfo {author}
  {\bibfnamefont {A.}~\bibnamefont {Goban}}, \bibinfo {author} {\bibfnamefont
  {D.}~\bibnamefont {Kedar}}, \bibinfo {author} {\bibfnamefont
  {C.}~\bibnamefont {Sanner}}, \bibinfo {author} {\bibfnamefont {J.~M.}\
  \bibnamefont {Robinson}}, \bibinfo {author} {\bibfnamefont {G.~E.}\
  \bibnamefont {Marti}},  \emph {et~al.},\ }\href {\doibase
  10.1038/s41566-019-0493-4} {\bibfield  {journal} {\bibinfo  {journal} {Nature
  Photonics}\ }\textbf {\bibinfo {volume} {13}},\ \bibinfo {pages} {714}
  (\bibinfo {year} {2019})}\BibitemShut {NoStop}%
\bibitem [{\citenamefont {Nakamura}\ \emph {et~al.}(2020)\citenamefont
  {Nakamura}, \citenamefont {Davila-Rodriguez}, \citenamefont {Leopardi},
  \citenamefont {Sherman}, \citenamefont {Fortier}, \citenamefont {Xie},
  \citenamefont {Campbell}, \citenamefont {McGrew}, \citenamefont {Zhang},
  \citenamefont {Hassan} \emph {et~al.}}]{Nakamura2020}%
  \BibitemOpen
  \bibfield  {author} {\bibinfo {author} {\bibfnamefont {T.}~\bibnamefont
  {Nakamura}}, \bibinfo {author} {\bibfnamefont {J.}~\bibnamefont
  {Davila-Rodriguez}}, \bibinfo {author} {\bibfnamefont {H.}~\bibnamefont
  {Leopardi}}, \bibinfo {author} {\bibfnamefont {J.~A.}\ \bibnamefont
  {Sherman}}, \bibinfo {author} {\bibfnamefont {T.~M.}\ \bibnamefont
  {Fortier}}, \bibinfo {author} {\bibfnamefont {X.}~\bibnamefont {Xie}},
  \bibinfo {author} {\bibfnamefont {J.~C.}\ \bibnamefont {Campbell}}, \bibinfo
  {author} {\bibfnamefont {W.~F.}\ \bibnamefont {McGrew}}, \bibinfo {author}
  {\bibfnamefont {X.}~\bibnamefont {Zhang}}, \bibinfo {author} {\bibfnamefont
  {Y.~S.}\ \bibnamefont {Hassan}},  \emph {et~al.},\ }\href {\doibase
  10.1126/science.abb2473} {\bibfield  {journal} {\bibinfo  {journal}
  {Science}\ }\textbf {\bibinfo {volume} {368}},\ \bibinfo {pages} {889}
  (\bibinfo {year} {2020})}\BibitemShut {NoStop}%
\bibitem [{\citenamefont {Brewer}\ \emph {et~al.}(2019)\citenamefont {Brewer},
  \citenamefont {Chen}, \citenamefont {Hankin}, \citenamefont {Clements},
  \citenamefont {Chou}, \citenamefont {Wineland}, \citenamefont {Hume},\ and\
  \citenamefont {Leibrandt}}]{Brewer2019}%
  \BibitemOpen
  \bibfield  {author} {\bibinfo {author} {\bibfnamefont {S.~M.}\ \bibnamefont
  {Brewer}}, \bibinfo {author} {\bibfnamefont {J.-S.}\ \bibnamefont {Chen}},
  \bibinfo {author} {\bibfnamefont {A.~M.}\ \bibnamefont {Hankin}}, \bibinfo
  {author} {\bibfnamefont {E.~R.}\ \bibnamefont {Clements}}, \bibinfo {author}
  {\bibfnamefont {C.-W.}\ \bibnamefont {Chou}}, \bibinfo {author}
  {\bibfnamefont {D.~J.}\ \bibnamefont {Wineland}}, \bibinfo {author}
  {\bibfnamefont {D.~B.}\ \bibnamefont {Hume}}, \ and\ \bibinfo {author}
  {\bibfnamefont {D.~R.}\ \bibnamefont {Leibrandt}},\ }\href@noop {} {\bibfield
   {journal} {\bibinfo  {journal} {Physical review letters}\ }\textbf {\bibinfo
  {volume} {123}},\ \bibinfo {pages} {033201} (\bibinfo {year}
  {2019})}\BibitemShut {NoStop}%
\end{thebibliography}%

\end{document}